\documentclass[twocolumn,showpacs,amsmath,amssymb]{revtex4}

\begin{document}
\title{Hydrodynamical thermotransport relaxation times of semiconductor
electrons via acoustic phonons}

\author{A.\ Rangel-Huerta$^+$ and M.A.\ Rodr\'{\i}guez-Meza$^*$}
\address{$^+$Facultad de Ciencias de la Computaci\'on
Benem\'{e}rita Universidad Aut\'{o}noma de Puebla,
14 Sur y San Claudio C.U., Puebla 72570 Mexico \\
$^*$Depto. de F\'{\i}sica, Instituto Nacional de Investigaciones
Nucleares, Apdo.\ Postal 18-1027, M\'{e}xico D.F.\ 11801, M\'{e}xico \\
e-mail: mar@nuclear.inin.mx}

\date{\today}

\begin{abstract}
We propose a hydrodynamic model to study the thermotransport properties
of semiconductor electrons. From the semiclassical Boltzmann equation
a set of balance equations is built for the relevant fields.
The electron density, the electron energy density, the electric current density and 
the heat flux
density are considered as the basic fields of direct transport and cross
effect fluxes. The kinetic relaxation times of the production terms are
calculated by considering the electron-acoustic phonon interaction as the
leading microscopic scattering process. To justify the long time
thermalization regime, the Onsager symmetry relations are proved, both on the
completely degenerate and non degenerate limits.
\end{abstract}

\pacs{05.20.Dd, 72.10.Bg,  72.15.Jf,  72.15.Lh \vspace{-4pt}}

\maketitle


\section{Introduction}

Hydrodynamical \ models have been widely used to describe thermotransport 
properties of
electrons in semiconductor materials. Complementary contributions of charge
transport and thermal dissipation fluxes can be included in the same model,
but due to the complexity of the system, and the amount of microscopic
processes involved, it is mandatory to separate the analysis of the
microscopic scattering processes of electrons.

As usual, a generalized hydrodynamic model of semiconductors requires of
constructing a hierarchy of balance equations for the complete set of
basic fields from the semiclassical Boltzmann transport 
equation (BTE)\cite{RaVe2002}. 
This hydrodynamic scheme constitutes a 
closed set of
balance equations for the relevant variables once the higher order moments
and the production terms have been completely evaluated.

Depending on the order of the approximation, an expansion of the non
equilibrium distribution function (NEDF) must be used to calculate the higher order
moments by means of the local variables. For example, in degenerate
semiconductors, a local Fermi-Dirac distribution function is used as a weight
function to take in to account the non equilibrium regime. Most of the
complications appear when calculating the production terms
corresponding to the non conserved variables. 
The largest difficulty arises as a result of
the intrinsic complexity 
of the collisional kernel of the BTE,
enhanced by the microscopic semiconductor scattering processes of electrons
which involve a variety of crystal lattice perturbations.

Due to all these considerations the closure problem is usually
solved by including some phenomenological components, such as the
relaxation time approximation, or the computational Monte Carlo analysis.
The result is a hybrid scheme of thermotransport of electrons between the
kinetic and phenomenological models (or kinetic and numerical Monte Carlo
method)\cite{AnTro1997,AnRoRu2000}. Following these arguments, we can
understand why most of the hydrodynamical models of thermotransport in
semiconductors have some approximations in the region of quasi-stationary
macroscopic fluxes. 
In this region, the perturbations drive the system not to far from the
local thermodynamic equilibrium, so it is possible to develop a complete
kinetic theoretical model considering a specific electron-lattice interaction
dominating the thermalization process.

In this work, a kinetic theory approach is considered to solve the
semiclassical BTE for the electrons in a semiconductors. The method of moments
derives into a set of balance equations up to the 8-fields approximation. The
basic fields of the electron density, electric current density, energy density,
and heat flux are the local relevant variables. The closure of the balance
equations is achieved by using the maximum entropy principle with a
NEDF, which allows to evaluate the
production terms without any phenomenological assumption, as is commonly done
in the literature\cite{RaVe2002,RoCaRa2003}. The electron-acoustic phonon
interaction is the leading scattering mechanism, i.e., the lattice
temperature is relatively low. The electric applied field is absent,
however, an internal electric field is generated because of the
inhomogeneties of the electron's distribution. The resulting kinetic
relaxation times are used to justify the linear hydrodynamic model by
checking the corresponding Onsager symmetry relations. This assures that
the linear response theory corresponds to the relaxation time approximation.


\section{The kinetic model}
We assume that a macroscopic state of the electrons is characterized by the eight
local relevant variables corresponding to the number density of the electron's 
charge, $n(\mathbf{x},t),$ which defines the mass density through 
$\rho (\mathbf{x},t)=m\,n(\mathbf{x},t)$, where $m$ is the electron's effective mass. The
electric current density, 
$J_{i}(\mathbf{x},t)=-en(\mathbf{x},t)\;v_{i}(\mathbf{x},t)$, 
where $v_{i}(\mathbf{x},t)$ is the hydrodynamic velocity and 
$e$ is the electrical charge of electrons. The energy density, 
$W(\mathbf{x},t)$, is defined as the average of 
$\mathcal{E=}$ $\frac{1}{2}m\,C_{\mathbf{k}}^{2}$. 
Here, 
$\mathbf{C}_{\mathbf{k}}$ is the peculiar velocity with components 
$C_{i}(\mathbf{x},\mathbf{k},t)=c_{i}(\mathbf{k})-v_{i}(\mathbf{x},t)$ and
$c_{i}(\mathbf{k})=\hbar k_{i}/m$ is the
velocity of electrons. To complete the set of basic fields, we have denoted
by $q_{i}(\mathbf{x},t)$ the local heat flux density.

The balance equations for describing the fluxes of electron's charge and
energy, according to Refs.\ \cite{RaVe2002,RaVeRo2005}, are:
%
\[
\frac{\partial \rho }{\partial t}-\frac{m}{e}\frac{\partial J_{i}}{\partial
x_{i}}=0,\text{ \ \ \ \ \ \ \ \ \ \ \ \ \ } 
\]
\[
\frac{\partial \rho W}{\partial t}+\frac{\partial q_{i}}{\partial x_{i}}-\mathcal{E}_{i}J_{i}=0, 
\]
\begin{equation}
\frac{\partial J_{i}}{\partial t}-\frac{eK_{B}}{m^{2}}\nu 
T^{\frac{3}{2}}\left( \frac{5}{3}I_{4}-\frac{\varphi }{K_{B}T}I_{2}\right) 
\frac{\partial T}{\partial x_{i}}
\end{equation}
\[
\text{ \ \ \ \ \ \ \ \ \ \ \ \ }-\frac{e^{2}}{m^{2}}\nu 
T^{\frac{3}{2}}I_{2}\mathcal{E}_{i}=
-\left( \frac{J_{i}}{\tau _{J}}+\frac{e}{K_{B}T}\frac{q_{i}}{\tau _{JQ}}\right) , 
\]
\[
\frac{\partial q_{i}}{\partial t}+\frac{5K_{B}^{2}}{3m^{2}}\nu 
T^{\frac{5}{2}}\left( \frac{7}{5}I_{6}-\frac{\varphi }{K_{B}T}I_{4}\right) 
\frac{\partial T}{\partial x_{i}} 
\]
\[
\text{ \ \ \ \ \ \ \ \ \ \ \ }+\frac{5eK_{B}}{3m^{2}}\nu 
T^{\frac{5}{2}}I_{4}\mathcal{E}_{i}=-\left( \frac{q_{i}}{\tau _{Q}}
+\frac{K_{B}T}{e}\frac{J_{i}}{\tau _{QJ}}\right) . 
\]

The effective local electric field is defined by 
$\mathcal{E}_{i}=(1/e)\partial \varphi /\partial x_{i}$, where $\varphi$ is the local
chemical potential. We have difined the constants, $\nu
=2^{9/2}K_{B}^{3/2}m^{5/2}\pi /h^{3}$, $K_{B}$ the Boltzmann constant, $h$
the Planck constant, and denoted by $I_{n}=I_{n}(\varphi /K_{B}T)$ the
Fermi-Dirac integrals\cite{RoCaRa2003}. Summation over repeated indices has
been considered. Note that the production terms, rhs of Eqs.\ (1), contains
contributions to the cross effect fluxes, which are proportional to 
$\tau_{JQ}^{-1}$ and $\tau _{QJ}^{-1}$ respectively. They are necessary to compare
the constitutive relations of the thermotransport coefficients under a
Onsager symmetry criterion.

The closure of the set of generalized hydrodynamic evolution equations is
achieved by using a NEDF. This function must be an
approximate solution of the Boltzmann equation coming from the maximum
entropy principle, which is assumed to be of the form 
$f=\ f^{FD}(\mathcal{E})(1+\Phi )$. The term $\Phi $ 
can be written in terms of the basic
fields\cite{RaVe2002}, 
%
%
\begin{equation}
\Phi =(1-f_{\mathbf{k}}^{FD})\left( F_{i}^{(3)}J_{i}+G_{i}^{(3)}q_{i}\right) ,
\end{equation}
where,
%
%
\begin{equation}
F_{i}^{(3)}=
\frac{21m^{2}}{e\nu K_{B}T^{\frac{5}{2}}}I_{J}\left( \mathcal{E}\right)
C_{i}\left( \mathbf{k}\right) ,
\end{equation}
%
%
%
\begin{equation}
G_{i}^{(3)}=\frac{15m^{2}}{\nu K_{B}^{2}T^{\frac{7}{2}}}I_{Q}
\left( \mathcal{E}\right) C_{i}\left( \mathbf{k}\right) .
\end{equation}

The functions $I_{J}\left( \mathcal{E}\right) $ and 
$I_{Q}\left( \mathcal{E}\right) $ are defined through 
the Fermi integrals as follows\cite{RaVeRo2005},
%
%
%
\begin{equation}
I_{J}(\mathcal{E})=\frac{I_{6}
-\frac{5\mathcal{E}}{7K_{B}T}I_{4}}{25I_{4}^{2}-21I_{2}I_{6}}\; ,
\end{equation}
and
%
%
\begin{equation}
I_{Q}(\mathcal{E})=\frac{I_{4}
-\frac{3\mathcal{E}}{5K_{B}T}I_{2}}{25I_{4}^{2}-21I_{2}I_{6}}.
\end{equation}

In this semiclassical hydrodynamical model, the weight function 
$\ f^{FD}(\mathcal{E})$ is a local Fermi-Dirac distribution function. 
Also, in Ref.\ \cite{RaVeRo2005}
we can see that the relaxation times have the following general
form for the electric current density,
%
%
\begin{eqnarray}
\frac{1}{\tau _{J}} &=&\sum_{\pm }\frac{-em}{3\hbar }\int d\mathbf{k}
d\mathbf{k}^{\prime } S\left( \mathbf{k,k}^{\prime }\right) 
\left( \mathbf{k}^{\prime }-\mathbf{k}\right) \\
&&\left[ f_{\mathbf{k}}^{FD}(1-f_{\mathbf{k}}^{FD})(1-f_{\mathbf{k}^{\prime
}}^{FD})F_{i}^{(3)}\right.  \nonumber \\
&&\left. \text{ \ \ \ \ \ \ \ \ \ \ \ \ \ \ \ \ \ }-\ f_{\mathbf{k}}^{FD}
f_{\mathbf{k}^{\prime }}^{FD}(1-f_{\mathbf{k}^{\prime }}^{FD})F_{i}^{(3)\prime }\right]  ,
\nonumber
\end{eqnarray}
and for the heat flux density,
%
%
\begin{eqnarray}
\frac{1}{\tau _{Q}} &=&\sum_{\pm }\frac{m^{2}}{6\hbar }
\int d\mathbf{k}d\mathbf{k}^{\prime }S\left( \mathbf{k,k}^{\prime }\right) 
\left( \mathcal{E}^{\prime }\mathbf{k}^{\prime }-\mathcal{E}\mathbf{k}\right) \\
&&\left[ f_{\mathbf{k}}^{FD}(1-f_{\mathbf{k}}^{FD})(1-f_{\mathbf{k}^{\prime
}}^{FD})G_{i}^{(3)}\right.  \nonumber \\
&&\left. \text{ \ \ \ \ \ \ \ \ \ \ \ \ \ \ \ \ \ }-\ f_{\mathbf{k}}^{FD}
f_{\mathbf{k}^{\prime }}^{FD}
(1-f_{\mathbf{k}^{\prime }}^{FD})G_{i}^{(3)\prime }\right] .  \nonumber
\end{eqnarray}
Similar expressions have been found for the cross effect fluxes. The signs
under the summation symbol indicate absorption and emmision respectively.


\section{The relaxation times}
To evaluate the kinetic relaxation times, we must specify an
electron-phonon interaction according to the perturbations in the
crystal lattice. In the present case, we have considered 
interactions with acoustic phonons
as the leading mechanism of the collisional processes of electrons. As usual,
the rate of electron collisions with acoustic phonons is given by\cite{Hamaguchi2001},
%
%
\begin{equation}
S\left( \mathbf{k,k}^{\prime }\right) =\frac{2\pi }{\hbar }\left[ 
\frac{D_{ac}^{2}K_{B}T}{2L^{3}\rho v_{s}^{2}}\right] \delta 
\left( \mathcal{E-E}^{\prime }\right) 
\end{equation}
where $D_{ac}$ is the electron-acoustic phonons deformation potential,
and $v_{s}$ represents the velocity of sound. After a substitution of Eq.\
(9) into Eqs.\ (7-8) and considering that the electron energy $\mathcal{E}$
is the integration variable, we obtain general expressions for the
relaxation times. The corresponding relaxation time for the electric current
density is,
%
\begin{eqnarray}
\tau _{J}^{-1} &=&\frac{7}{2}\;\frac{\left( 2\pi \right) ^{2}D_{ac}^{2}
m^{\frac{3}{2}}}{\hbar ^{4}\rho v_{s}^{2}(K_{B}T)^{\frac{1}{2}}} \\
&&\int I_{J}(\mathcal{E})\left[ f^{FD}(\mathcal{E})\right] ^{2}\left(
1-f^{FD}(\mathcal{E})\right) \mathcal{E}^{2}d\mathcal{E}  \nonumber
\end{eqnarray}

A similar procedure gives rise to all the other relaxation time expressions.
The following are the expressions for $1/\tau _{Q}$, $1/\tau _{JQ}$, and 
$1/\tau _{QJ}$. For the heat flux we have, 
%
\begin{eqnarray}
\tau _{Q}^{-1} &=&\frac{5}{2}\;\frac{\left( 2\pi \right) ^{2}D_{ac}^{2}
m^{\frac{3}{2}}}{\hbar ^{4}\rho v_{s}^{2}(K_{B}T)^{\frac{3}{2}}} \\
&&\int I_{Q}(\mathcal{E})\left[ f^{FD}(\mathcal{E})\right] ^{2}\left(
1-f^{FD}(\mathcal{E})\right) \mathcal{E}^{3}d\mathcal{E}  \nonumber
\end{eqnarray}
and the crossed effect relaxation times, $\tau _{JQ}$ and $\tau _{QJ},$ are
given by,
%
%
\begin{eqnarray}
\tau _{JQ}^{-1} &=&\frac{5}{2}\;\frac{\left( 2\pi \right) ^{2}D_{ac}^{2}
m^{\frac{3}{2}}}{\hbar ^{4}\rho v_{s}^{2}(K_{B}T)^{\frac{1}{2}}} \\
&&\int I_{Q}(\mathcal{E})\left[ f^{FD}(\mathcal{E})\right] ^{2}\left(
1-f^{FD}(\mathcal{E})\right) \mathcal{E}^{2}d\mathcal{E}  \nonumber
\end{eqnarray}
%
%
%
\begin{eqnarray}
\tau _{QJ}^{-1} &=&\frac{7}{2}\;\frac{\left( 2\pi \right) ^{2}D_{ac}^{2}
m^{\frac{3}{2}}}{\hbar ^{4}\rho v_{s}^{2}(K_{B}T)^{\frac{3}{2}}} \\
&&\int I_{J}(\mathcal{E})\left[ f^{FD}(\mathcal{E})\right] ^{2}\left(
1-f^{FD}(\mathcal{E})\right) \mathcal{E}^{3}d\mathcal{E}\text{.}  \nonumber
\end{eqnarray}
By incorporating this expressions of the relaxation times in the hydrodynamical
set of transport equations we obtain a completely closed model containing
no free or phenomenological parameters. It is very important to note that
though the structure of the transport equations is given, and does not
depend on a microscopic model for the interaction, the specific values of
the relaxation times depend closely on the model of the electron-lattice
interaction.


\section{Onsager symmetry}
As is well known, the Onsager symmetry principle projects microscopic
reversibility of the interactions onto a macroscopic scale through the transport
symmetry relations. In this section we will show that, in the linear
response regime, the semiconductor transport of charge, under internal
electric fields variations and thermal gradients, can be characterized by a
set of constitutive relations with thermoelectric transport coefficients
displaying symmetry on the cross effect fluxes. This point deserves some
discussion because the microscopic interactions of the semiconductor
electrons are regulated by a process of creation and annihilation of phonons.
So there is no guarantee of microscopic reversibity. However, because of the
elastic collision between electrons and acoustic phonons, our result reflects
the existence of a kind of a more general detailed balance principle of
collisions.

We show the validity of the Onsager symmetry relations, at the end of the
thermalization process and both in the completely degenerate as well non
degenerate limits, by following a general expression that must be
satisfied (see Ref.\ \cite{RaVe2002}):
%
\begin{equation}
\tau _{J}\left( 5I_{4}-3I_{2}\frac{\tau _{Q}}{\tau _{QJ}}\right) =-\tau
_{Q}\left( 5I_{4}-7I_{6}\frac{\tau _{J}}{\tau _{JQ}}\right) .
\end{equation}%
First, we are considering the kinetic relaxation times in the complete
degenerate limit. In this case, integration with respect to energy, and
assuming the Fermi-Dirac distribution function satisfying 
$f^{FD}(\mathcal{E})\left( 1-f^{FD}(\mathcal{E})\right) 
=\delta \left( \mathcal{E-\varphi }\right)$, 
derives into the following expressions for the relaxation times:
%
%
%
\begin{eqnarray}
\tau _{J}^{-1} &=&\frac{7}{2}\;\upsilon \text{ }\frac{I_{6}
-\frac{5\varphi }{7K_{B}T}I_{4}}{25I_{4}^{2}-21I_{2}I_{6}}, \\
\tau _{Q}^{-1} &=&\frac{5}{2}\;\upsilon \text{ }\frac{I_{4}
-\frac{3\varphi }{5K_{B}T}I_{2}}{25I_{4}^{2}-21I_{2}I_{6}}\frac{\varphi }{K_{B}T},  
\nonumber
\\
\tau _{JQ}^{-1} &=&\frac{5}{2}\;\upsilon \text{ }\frac{I_{4}-\frac{3\varphi 
}{5K_{B}T}I_{2}}{25I_{4}^{2}-21I_{2}I_{6}},  \nonumber \\
\tau _{QJ}^{-1} &=&\frac{7}{2}\;\text{ }\upsilon \text{ }\frac{I_{6}
-\frac{5\varphi }{7K_{B}T}I_{4}}{25I_{4}^{2}-21I_{2}I_{6}}\frac{\varphi }{K_{B}T}.
\nonumber
\end{eqnarray}
Here, we have introduced the parameter $\upsilon ,$ which depends on the
chemical potential of electrons and the local temperature,
%
%
\begin{equation}
\upsilon =\left( \frac{2\pi mD_{ac}\varphi }{K_{B}T}\right) ^{2}
\sqrt{\frac{K_{B}T}{m}\text{ }}\frac{K_{B}T}{\hbar ^{4}\rho v_{s}^{2}}.
\end{equation}
A direct substitution of these expressions into the Onsager conditions show
us that they are well satisfied. On the non degenerate case, the Onsager
symmetry relation is well satisfied too. If we consider this limit, 
$f^{FD}(\mathcal{E})\ll 1$ and it becomes a Maxwellian distribution function.
Under these conditions the relaxation times can be expressed as follows,
%
\begin{eqnarray}
\tau _{J}^{-1} &=&\frac{1}{4}\;\nu _{1}\text{ }
\left( \frac{K_{B}T}{2}\right) ^{3}\text{ }2!\frac{\left( 14I_{6}-15I_{4}\right) }{\left(
25I_{4}^{2}-21I_{2}I_{6}\right) } \\
\tau _{Q}^{-1} &=&\frac{1}{2}\;\nu _{3}\text{ }
\left( \frac{K_{B}T}{2}\right) ^{4}\text{ }3!\frac{\left( 5I_{4}-6I_{2}\right) }{\left(
25I_{4}^{2}-21I_{2}I_{6}\right) }  \nonumber \\
\tau _{JQ}^{-1} &=&\frac{1}{2}\;\nu _{1}\text{ }
\left( \frac{K_{B}T}{2}\right) ^{3}\text{ }1!\frac{\left( 10I_{4}-9I_{2}\right) }{\left(
25I_{4}^{2}-21I_{2}I_{6}\right) }  \nonumber \\
\tau _{QJ}^{-1} &=&\frac{1}{2}\;\nu _{3}\text{ }
\left( \frac{K_{B}T}{2}\right) ^{4}\text{ }3!\frac{\left( 7I_{6}-10I_{4}\right) }{\left(
25I_{4}^{2}-21I_{2}I_{6}\right) }  \nonumber
\end{eqnarray}%
here we have defined the parameters,
%
\begin{equation}
\nu _{n}=\frac{\left( 2\pi D_{ac}\right) ^{2}m^{\frac{3}{2}}}{\hbar ^{4}\rho
v_{s}^{2}\left( K_{B}T\right) ^{\frac{n}{2}}}\exp 
\left( \frac{2\varphi }{K_{B}T}\right) ,\text{ \ }n=1,2,...
\end{equation}

Once again, the Onsager condition is well satisfied in this case. Next, what
we can see is if the full integral expressions for the kinetic relaxation
times satisfy the symmetry condition. However, this requires a
computational analysis and a correct interpretation of the numerical
uncertainties. We can speculate that the Onsager symmetry is going to be
well satisfied in all the interval of the temperature varations, because of
the regime in which we are working (the last stage of thermalization)
and the assumption of the microscopic detailed balance under elastic
electron-phonon collision being satisfied.

The general condition of symmetry, in Eq.\ (14), also appears in 
Refs.\ \cite{RaVe2002,RaVeRo2005}, among others, with respect to 
the Onsager symmetry
relations of the thermotransport coefficients. In particular in 
Ref.\ \cite{RaVeRo2005} it is shown that for thermotransport in polar semiconductors,
where the main microscopic interaction mechanism is the electron-polar
optical phonons, the relation specifies that only some values of $\varphi $
are satisfying the symmetry condition, i.e., only certain values are allowed
for the chemical potential in comparison to the optical phonon energy.
Though this condition seems a very restrictive one, we must note that the
model corresponds only to the behavior of electron population. The phonons
become the reservoir of lattice perturbations, so they are not taken into
account in order to preserve microscopic reversibility\cite{CaRo1991}. Hence the
Onsager symmetry relation depends strongly on the thermalization regime
where a different microscopic electron-lattice interactions may be dominating
the relaxation processes.


\section{Discussion and  conclusions}
We have developed a kinetic theoretical approach to describe a generalized
hydrodynamic behavior of semiconductor electrons at the end of the
thermalization process. The semiclassical Boltzmann transport equation have
been solved up to the eight moments method approximation in such a way that
the heat flux density is considered as a relevant variable in the resulting
hydrodynamical model. The closure of the balance equations is achieved by
evaluating the higher order moments, as well as the production terms,
through a non equilibrium distribution function according to the maximum
entropy principle as has been described in literature. 
In case that
the electron-acoustic phonon scattering is the leading microscopic
interaction,
the production terms
can be expressed by means of a set of kinetic relaxation times.

As a consequence of the theoretical model developed in this paper,
we have shown analytically that condition (14) is satisfied
in both degenerate and non degenerate limits. This means that
Onsager symmetry relations are fulfilled. Even more, 
and given that for the electron-acoustic phonon scattering the microscopic
reversibilty is valid, we believe that the general expressions for
the relaxation times, Eqs.\ (10)--(13), will give that Onsager symmetry
relations are also satisfied under more general conditions.

In general, the electron transport properties in semiconductors depend
on the scattering processes that electrons suffer. In addition, these 
transport processes can
be expressed in terms of the relaxation times for each scattering mechanism.
Relaxation times are usually obtained by Monte Carlo 
Simulations\cite{Anile1995}.
For the case of electron-acoustic phonon interaction we have found 
closed
general
expressions, Eqs.\ (10)--(13), which together with the general expressions
found in Ref.\ \cite{RaVeRo2005} for electron-optical phonon interaction
provide us with a theoretical framework to describe the thermotransport processes
of electrons in semiconductors under more general conditions with no free
or phenomenological parameters. A complete scheme will be obtained when
we include electron-intervalley phonons and electron-impurity scattering processes.
The case of piezoelectric phonon perturbations is rather direct and follows the same
argument as acoustic phonon perturbations.
A detailed analysis of the relaxation
times as function of lattice temperature and electron concentration will give
us useful information of the main relaxation channel under the imposed 
conditions\cite{Gantmakher1987,CaRo1991}.
For example, in Ref. \cite{Anile1995} it is shown that, for a silicon device,
the energy relaxation time (which is the characteristic time in which the
electron energy relaxes to the lattice) is substantially larger than the
relaxation times for momentum and heat flux. The situation could be
very different for Bulk GaAs. For high temperatures (300 K) and low 
electron concentration (of the order or less than $10^{16}$ cm$^{-3}$) 
the main collision mechanism is electron-acoustic interactions. Whereas,
for low temperatures and high electron concentration where screening
effect should be taken into account, the electron-optical phonon interaction
should be the main ralaxation mechanism\cite{CaRo1991}.

\end{document}